# Generalized Statistical Models of Voids and Hierarchical Structure in Cosmology

Aram Z. Mekjian

Rutgers University, Department of Physics and Astronomy, Piscataway, NJ. 08854
&
California Institute of Technology, Kellogg Radiation Lab 106-38, Pasadena, Ca 91125

**ABSTRACT**

Generalized statistical models of voids and hierarchical structure in cosmology are developed. The often quoted negative binomial model and frequently used thermodynamic model are shown to be special cases of a more general distribution which contains a parameter $a$. The parameter $a$ is related to the Levy index $\alpha$ and the Fisher critical exponent $\tau$, the latter describing the power law fall off of clumps of matter around a phase transition. The parameter $a$, exponent $\tau$, or index $\alpha$ can be obtained from properties of a void scaling function. A stochastic probability variable $p$ is introduced into a statistical model which represent the adhesive growth of galaxy structure. For $p < 1/2$, the galaxy count distribution decays exponential fast with size. For $p > 1/2$, an adhesive growth can go on indefinitely thereby forming an infinite supercluster. At $p = 1/2$ a scale free power law distribution for the galaxy count distribution is present. The stochastic description also leads to consequences that have some parallels with cosmic string results, percolation theory and phase transitions.

*Subject headings*: cosmology: large scale structure – galaxies: clustering, hierarchical structure & voids - general: statistical models, Levy distributions, cosmic strings, percolation theory, phase transitions

## 1. INTRODUCTION

The distribution of galaxies on large scales shows geometrical features such as walls, filaments and voids. Understanding these features and related issues of clustering and density fluctuations in a gravitational many body system is a major endeavor in cosmology. Theories of the origins of large scale structure and anisotropy of the cosmic microwave background has proceeded along several lines. One approach is based on an amplification of quantum fluctuations during inflation. The other main approach is based on symmetry breaking during a phase transition in the early universe which plays an important role in its evolution in time. Associated with this phase transition are topological defects such as cosmic strings, domain walls, and texture. Loops of cosmic string may be centers for the accretion of galactic matter and be a source of a clumping seen in the distribution of matter. The clustering of galaxies depends on many features which involve the cosmological parameters, the dark matter distribution and the coupling and evolution of this component to the luminous component. Another, less dynamical and more phenomenological approach has involved studies of the properties of the distribution of galaxies and the large scale structure through correlation studies. In this approach, the analysis focuses is on the nature of the correlations looking for patterns that may give physical insight into the processes that are responsible for the structure. One pattern is that of an hierarchical structure. Early methods of analysis of clustering of galaxies were based on 2-point correlation functions (Peebles 1980), which offered a lowest order correlation approach to such clustering. A more recent investigation of the importance of the two point correlation function at large scales was carried out by Durrer et. al. 2003. Methods for studying correlations of higher order have also been developed. The simplest assumption is that higher order correlations obey an hierarchy of scaling where higher order correlations are related to lower order two body correlations. Hierarchical correlations were shown to appear in non-linear regimes (Peebles 1980). They are commonly discussed by many others and a recent review with references can be found in Bernardeau *et al.* (2002). The void probability functions VPF offers a useful way of understanding some of the higher order correlation features. Recent observations from the 2dF Galaxy Redshift Survey by Croton *et al.* (2004) show that a reduced void measure is a useful way of investigating the large scale structure seen and that the results are in excellent agreement with a paradigm of hierarchical scaling. Studies of the physical properties of voids have also been studied by Hoyle and Vogeley (2004).

   Distributions taken from quantum optics and photon count models (Carruthers and Shih 1983)] have been somewhat successfully applied to the analysis of the large scale structure seen. One of the most familiar of these distributions is based on a negative binomial probability description. The negative binomial model implies a specific hierarchical structure to the distribution which will be discussed. The negative binomial distribution has also been used in particle physics phenomenology and, in particular, in understanding intermittency phenomena. The recent analysis of cosmological data by Croton *et al.* (2004) has shown small departures from the negative binomial result for the VPF. The VPF associated with other distributions were also considered by them. These included a Gaussian model, a thermodynamic model, a lognormal distribution and a distribution based on a BBGKY distribution. The negative binomial best characterized the data amongst the specific distribution considered.

   In previous papers (Mekjian 2001 and also Lee and Mekjian 2004) and in a different context (particle and nuclear physics, count probabilities in quantum optics), a generalization of the negative binomial model was developed using a theoretical framework based on statistical mechanics. This generalization involved the introduction of a parameter, called $a$, into a statistical model of count probabilities and nuclear and particle cluster distributions. The formation of clusters in an expanding many body system of hadronic matter has some aspects that are similar to the clustering of galaxies in an expanding universe mentioned above. The statistical model also has some similarities with droplet models of phase transitions around a critical point. The parameter $a$ is analogous to the Fisher critical exponent $\tau$ (Stanley 1971)). The exponent $\tau$ describes the power law fall off of droplet sizes associated with the phenomena of critical opalescence occurring at the critical point of a second order phase transition. The parameter $a$ can be related to the Levy index $\alpha$ -Levy (1954), Feller (1971), Sato (1999). The importance of the Levy distribution has been stressed by Mandelbrot (1982). In the context of particle physics and quantum optics, the introduction of the parameter $a$ enabled one to connect the negative binomial model ($a = 1$) to a second model ($a = 1/2$) used in particle physics and in quantum optics and also in descriptions of galaxy clustering. Specifically, in the present context of galaxy clustering, the limiting case of $a = 1/2$ has the same void distribution as the thermodynamic model initially due to Saslaw and Hamilton (1984) and modified by Fry (1986). Fry also considered several other possibilities such as an hierarchical Poisson model and an hierarchical *BBGKY* model. Moreover, in particle physics, the distribution is also connected to a Feynman/Wilson gas (Mekjian 2001) and in quantum optics the count distribution was initially due to Glauber (1963) and also discussed in further detail in Klauder and Sudarshan (1968). The result arises in photon distributions from field emission from states with Lorentzian line shapes. The negative binomial and this second distribution have somewhat different count distributions, void scaling and hierarchical structure properties. Generalizations of the negative binomial distribution in particle phenomenology have also been developed by Hegyi (1993, 1996) and related to the Fox function.

A review of the importance of various scaling laws in the distribution of galaxies can be found in Jones *et al.* (2005). These authors considered the power law behavior of the two point correlation function, both radial and angular, higher order correlation functions such as the three point correlation function, the hierarchical structure of the $q$-order correlations, the power spectrum, counts-in-cell and the associated void probability function, fractal and multi-fractal measures. Scaling features of voids and their connection to fractal models have been studied by Gaite and collaborators (2002, 2005, 2006). Studies of the scaling of moments of counts distribution with cell size $L$ have been proposed as a method of obtaining the scaling dimension of the system (Martinez and Saar 2002, Borgani 1995). A discussion of the three point function is given in Taka and Jain 2003. A recent search for voids can be found in Hoyle and Vogeley 2004.

## 2. CORRELATION FUNCTIONS, HIERARCHICAL STRUCTURE AND VOID PROBABILITY

The probability of finding $N$ galaxies in a selected volume $V$ randomly placed gives the count probability $P_N$. The void probability is the special case $N = 0$. The $P_0$ is connected to mean $k$-point correlation function $\xi_k$ (White 1979) through the relation

$$P_0 = \exp[-\sum_{k=1}^{\infty} \frac{(-<N>)^k}{k!} \xi_k ] \tag{1}$$

The $<N>$ is the mean number of galaxies in $V$. To obtain $P_0$ from $\xi_k$, all correlation moments $\xi_k$ have to be determined. The relation simplifies in hierarchical models where the $k$-point correlation function can be expressed in terms of the 2-point correlation function through a connection that is expressed as

$$\xi_k = A_k \xi_2^{k-1} \tag{2}$$

Such a organization in the $\xi_k$'s is referred to as hierarchical structure. The $A_k$ is a scaling coefficient that is determined by the hierarchical model. Using this relation, the term

$(-<N>)^k \xi_k$ in the expression for the void probability simplifies to $<N>(<N>\xi_2)^{k-1}$. A quantity $\chi \equiv -Ln(P_0)/<N>$ called the reduced void probability function now has the feature of being a function of the specific combination $<N>\xi_2$ only. Namely

$$\chi = \sum_{k=1}^{\infty} \frac{A_k}{k!}(<N>\xi_2)^{k-1} \qquad (3)$$

The behavior of $\chi$ with $<N>\xi_2$ depends on the scaling amplitudes $A_k$. For example, if $A_k = (k-1)!$, the sum over $k$ is simply

$$\chi = \sum_{k=1}^{\infty} \frac{1}{k}(<N>\xi_2)^{k-1} = \frac{1}{<N>\xi_2} Ln(1+<N>\xi_2) \qquad (4)$$

This particular behavior of $\chi$ as a function of $<N>\xi_2$ arises from a negative binomial distribution as discussed in the next section. The $\xi_2$ and the fluctuation are related by $<N^2>-<N>^2 = <N>+\xi_2<N>^2$ or through the factorial moment result $<N(N-1)> = <N>^2(1+\xi_2)$.

Hierarchical pictures have been usefully applied in understanding the large scale structure from a dynamical or evolutionary view. Several descriptions have been developed based on hierarchical scaling. For example, small scale clumps are formed in regions where the density fluctuations are large and are produced earlier than large scale clumps. The small scale clumps are subsequently captured by large scale clumps in a hierarchical clustering process. The Press-Schechter (1974) theory is a statistical picture of this clustering.

In the present paper, a theoretical framework is developed to describe the large scale structure in terms of a statistical model based on a canonical and grand canonical ensemble. The probability $P_N$ is obtained from a fugacity expansion of the grand canonical ensemble. Specific choices of the form of the grand canonical ensemble lead to particular probability distributions such as the negative binomial model and the thermodynamic model. A generalized choice based on a hypergeometric description (sect 4.8) contains these two frequently used cases as special limits. Sections 4.4, 4.5 and 4.6 discuss specific distributions. The void scaling and hierarchical structure of the hypergeometric description are developed in further detail in sections 5.1 and 5.2. Before presenting these results, a brief summary of some of the ideas to be developed is presented in the next section using the negative binomial results to establish some of the useful quantites to be considered. Further highlights of the results are as follows. Section 4.3 contains a clan variable description while section 4.7 introduces a stochastic probability variable $p$ into the theoretical framework. The $p$ can be viewed as an evolutionary variable describing the adhesive growth of galaxy clusters. Parallels and consequences of this picture are further discussed in sections 6&7. Sect. 6 outlines a formal parallel with a result from cosmic strings and sec. 7 has a discussion of a parallel with percolation theory and phase transitions.

## 3. COUNT DISTIBUTIONS AND PARALLELS WITH QUANTUM OPTICS

Probability count distributions appear in many area of physics. Of relevance to this paper is a count distribution taken from quantum optics and was subsequently was used as a phenomenological model of galaxy count distributions. Specifically, Carruthers and Shih (1983) used the negative binomial distribution for this purpose. This distribution has a probability

$$P_N = \binom{N+\kappa-1}{N} p^\kappa (1-p)^N =$$
$$\binom{N+\kappa-1}{N} \left(\frac{<N>/\kappa}{1+<N>/\kappa}\right)^N \left(\frac{1}{1+<N>/\kappa}\right)^\kappa \quad (5)$$

the probability $p$ determines the mean number $<N>$ and $\kappa$ is a parameter that is appears in the expression for the fluctuation $\delta N^2 \equiv <N^2> - <N>^2$. Namely,

$$p = \frac{1}{(1+<N>/\kappa)}, \quad <N> = \kappa \frac{1-p}{p}$$
$$<N^2> - <N>^2 = <N>(1+<N>/\kappa) = \kappa(1-p)/p^2 \quad (6)$$

Properties of the negative binomial are also written in terms of the two variables $\widehat{P}, \widehat{Q}$ defined by $\widehat{Q} \equiv 1/p, 1-p \equiv \widehat{P}/\widehat{Q}$ or $\widehat{P} = (1-p)/p$. Then $<N> = \kappa \widehat{P}$ and $\delta N^2 = \kappa \widehat{P} \widehat{Q}$. Clan variables, introduced by Van Hove (1989), have also been used to characterize the negative binomial distribution and other probability distributions. The clan variable representation for the negative binomial distribution and other infinitely divisible probability distributions are given below in a separate section-sect 4.3. The void scaling function for the negative binomial is

$$\chi_{NB} = \frac{1}{\xi_2 <N>} Ln(1+\xi_2 <N>) \quad (7)$$

The negative binomial result $\chi_{NB}$ will be shown to be a special case of a more general void distribution which involves a parameter called $a$. The parameter $a$ will be related to the Levy stable index $\alpha$ and the Fisher critical exponent $\tau$ in section 4.8. The distribution to be developed is based on a generalized statistical model which has a void scaling distribution $\chi_a$ that is

$$\chi_a = \frac{1}{(1-a)(\xi_2 <N>/a)}\left[\left(1+\frac{\xi_2 <N>}{a}\right)^{1-a} - 1\right] \tag{8}$$

The negative binomial limit is $a \to 1$ limit of $\chi_a$. An often quoted void distribution is that of a model referred to as the thermodynamic model (Saslaw and Hamilton (1984), Fry (1986)) which has the special value $a = 1/2$.

It is useful to compare the negative binomial distribution with another frequently occurring distribution which is the count probability of Poisson statistics which has a probability distribution $P_N$ and variance $\delta N^2$ given by:

$$P_N = \frac{<N>^N}{N!}\exp(-<N>)$$
$$\delta N^2 \equiv <N> \tag{9}$$

Thus, in the Poisson case $\xi_2 = 0$ and consequently the Poisson void scaling function $\chi_P = 1$ for all $\xi_2 <N>$. In galaxy count distributions, the Poisson result represents the extreme limit of VPF being independent of the scaling variable $\xi_2 <N>$. By comparison, the importance of the Poisson distribution in quantum optics is in its connection to coherent states. A Poisson distribution represents a coherent signal while the negative binomial distribution is a chaotic distribution from noise. The negative binomial distribution is a generalization of the Planck distribution which has the index $\kappa = 1$. The Planck distribution gives a statistical enhancement in the occupancy of a quantum level arising from Bose-Einstein statistics. In phase space, bosons will clump together because of this statistical attraction leaving under populated regions or void-like regions in this phase space. This can be contrasted with a Poisson distribution with a random distribution of points lacking such correlations. Distributions based on combinations of Poisson and negative binomial are also discussed in sect 4.5 and offer the possibility of studying the interplay of these two distributions and its effect on void distributions. A discussion of a combination of a Poisson distribution and the thermodynamic model distribution is developed in sect. 7 where a parallel with percolation theory is also given.

## 4. A STATISTICAL MECHANICS APPROACH

### 4.1 *General considerations*

The connection of probability distributions with statistical mechanics can be seen by considering the connection of generating functions for probability distributions and the grand canonical ensemble as a generating function of the canonical ensemble of a fixed number of particles $N$. The generating function for a probability distribution can be written as

$$\exp(\sum_{k=1}^{\infty} C_k (u^k - 1)) = \sum_{N=0}^{\infty} P_N u^N \tag{10}$$

The extra $-1$ in the exponent of this equation is the normalization factor necessary so that $\sum_{N=0}^{\infty} P_N = 1$. The exponential factor $\exp(\sum_{k=1}^{\infty} C_k u^k) \equiv Z_{gen}(\vec{C}, u) = \sum_{N=0}^{\infty} Q_N u^N$ generates a set of functions $Q_N$, $N = 0,1,2,...,\infty$, such that $P_N = Q_N / Z_{gen}(\vec{C}, u = 1)$. The vector $\vec{C} \equiv \{C_1, C_2,..\}$. In particle multiplicity distributions, the $C_k$ are sometimes referred to as combinants (Gyulassy, Kauffman, and Wilson 1978). The $C_k$ for Poisson statistics is $C_k = C_1 \delta_{k,1}$ so that only $C_1 \neq 0$, with $C_1 = <N>$. For the negative binomial distribution, the $C_k = xt^k / k$ has a distribution obtained from a logarithmic expansion of $-xLn(1-t)$. The $x, t$ determine the mean and fluctuation through

$$<N> = x \frac{t}{1-t}$$
$$<N^2> - <N>^2 = <N> (1 + \frac{<N>}{x}) \tag{11}$$

Comparing this last result with eq.6, the following identifications can be made: $x \leftrightarrow \kappa$ and $t \leftrightarrow 1 - p$.

A strong similarity exists between the generating function for $P_N$ and a statistical mechanics connection relating the canonical ensemble partition function $Z_N$ and grand canonical ensemble partition function $Z_{gc}$ which reads

$$Z_{gc}(\beta, \mu) = \sum_{N=0}^{\infty} Z_N u^N = \sum_{N=0}^{\infty} Z_N \exp(\beta \mu N) \tag{12}$$

The $u = \exp(\beta \mu)$ is the fugacity, the $\mu$ is the chemical potential and the $\beta = 1/T$, where the Boltzmann constant $k_B = 1$. The $Z_0 = 1$ and corresponds to a system with no particles. The $Z_0$ is related to the void probability when discussing void scaling in multiplicity distributions. This parallel between the grand canonical partition function and the generating function of a probability distribution is an essential point that will be used in this paper. Instead of using the combinant notation $C_k$, the notation $x_k$ will be used in the statistical model. The probability that a system has $N$ particles is simple the ratio

$$P_N = \frac{Z_N}{Z_{gc}(\beta, \mu)_{\mu \to 0}} = \frac{Z_N}{\sum_{N=0}^{\infty} Z_N} \tag{13}$$

The void probability $P_0$ follows from

$$P_0 = \frac{1}{Z_{gc}(\beta,\mu)_{\mu\to 0}} = \frac{1}{\sum_{N=0}^{\infty} Z_N} \quad (14)$$

4.2 *Examples and motivations from statistical mechanics*

Some examples will now be given to establish some of the methods to be used. For example, a system with $Z_N = x^N/N!$, has the $\Sigma Z_N = \exp(x)$. A Poisson probability distribution for $P_N$ arises since

$$P_N = \frac{1}{N!} x^N \exp(-x) = \frac{1}{N!} <N>^N \exp(-<N>) \quad (15)$$

The last equality follows from the result that $x = <N>$. An example of a partition function that has the behavior $Z_N = x^N/N!$ is that of a non-degenerate ideal gas of particles. The one particle partition function is $Z_1 = x = g_S \int V d^3 p \exp(-\vec{p}^2/2mT)/h^3 = g_S V/\lambda_T^3$. The $Z_N = (Z_1)^N/N!$ and the $Z_N$ satisfies a one term recurrence relation $Z_N = xZ_{N-1}/N$. The $V$ is the volume and $g_S$ is the spin degeneracy factor $=2s+1$ of the particle. The $\lambda_T^3$ is the quantum volume given by $\lambda_T^3 = h^3/(2\pi mT)^{3/2} = h^3/(2\pi m/\beta)^{3/2}$. The analog of $C_k$ is now $x_k$ and for this case $x_k = x\delta_{1,k}$ so that only $k=1$ exists. The grand canonical ensemble for this ideal case is $Z_{gc}(\beta,\mu) = \exp(xu) = \exp(x\exp(\beta\mu))$. The chemical potential $\mu$ of the grand canonical ensemble is obtained by specifying the $<N>$ in the system: $\mu = T \ln(<N> \lambda_T^3/V)$. The Poisson distribution is the high temperature Maxwell Boltzmann limit of ideal gasses. Non-Poissonian distribution arise when particles are indistinguishable and the system has some degree of degeneracy as in Bose-Einstein or Fermi-Dirac distributions. Non-interacting ideal Bose-Einstein gases (spin 0,1,2,…) have (Feynman, R.P., 1978) $x_k = x/k^{5/2} + 1/k$ while non-interacting ideal Fermi-Dirac gases (spin 1/2,3/2,..) have $x_k = (-1)^{k+1} x/k^{5/2}$. These results depend on the number or dimensions $d$ in the system which was taken as 3. For $d$ dimensions the factor $x/k^{5/2} \to (L^d/\lambda_T^d)/k^{1+d/2}$ with $L$ the length of one side of $V$. The exponent of $k$, which is $1+d/2$, determines the rate of fall off of $x_k$ with $k$. For these specific cases of non-interacting Fermi-Dirac and Bose-Einstein gases, the exponent is well known. For interacting systems near a second order phase transition, critical exponents appear and an exponent that describes the fall off of cluster yields in a liquid/gas phase transition is the Fisher exponent $\tau$ (Stanley 1971).

Once $x_k$ is specified, the canonical partitions functions can be obtained from a

recurrence relation (Chase and Mekjian 1995) for the cases considered which involves the sum:

$$Z_N = \frac{1}{N} \sum_{N=0}^{N} k x_k Z_{N-k} \qquad (16)$$

with $Z_0 = 1$. This result is an extension of the previous non-interacting ideal gas result $Z_N = x Z_{N-1} / N$. The ideal gas result follows when $x_1 = 0, x_k \neq 0, k = 2,3,...$ . The $Z_{gc}$ can be shown to be (Mekjian 1990)

$$Z_{gc} = \exp(\sum_{k=1}^{\infty} x_k u^k) \qquad (17)$$

The canonical ensemble $Z_N$ for any $x_k$ whose grand canonical ensemble is given by eq.17 also has another representation in terms of a weight over the partition of $N$. The weight over all partitions is the main theoretical framework of statistical ensembles. Specifically, consider the partitions of $N$ into smaller groups or clusters of size k, with $n_k$ subgroups or clusters of size $k$. This procedure defines partition of $N$. The partition can be specified by given a particular set $(n_1, n_2,..., n_N) \equiv \vec{n}$ with the constraint $N = \Sigma k n_k$. A weight $W_N(\vec{n}, \vec{x})$, where $\vec{x} \equiv (x_1, x_2,..., x_N)$ is given to each $\vec{n}$ which has a structure

$$W_N(\vec{n}, \vec{x}) = \prod_{k=1}^{N} \frac{x_k^{n_k}}{n_k!} \qquad (18)$$

The $x_k$ will contain the underlying physical quantities. Examples of $x_k$ will be given shortly. The $n_k!$ are Gibbs factorials. The $Z_N$ is determined by summing $W_N(\vec{n}, \vec{x})$ over all event-by-event histories or possibilities of the vector $\vec{n}$:

$$Z_N = \sum_{\vec{n}} \prod_{k=1}^{N} \frac{x_k^{n_k}}{n_k!} \qquad (19)$$

This particular structure of the weigth of eq.18 leads to the recurrence relation of eq.16 and the form of eq.17 for $Z_{GC}$. A further significance to $x_k$ can be developed by considering clan variables.

4.3 *Clan variables*

Clan variables, introduced by Van Hove (1986), have also been used to characterize the negative binomial distribution and other probability distributions. The clan variables $N_C, n_C$ are defined as follows. The number of clans (in the notation used in this paper) is $N_C \equiv Ln(Z_{GC}(\vec{x}, u = 1)) = \Sigma x_k$, while the mean number of particles per clan is just the ratio $n_C \equiv <N>/N_C = \Sigma k x_k / \Sigma x_k$. The $N_C$ is the zeroth moment of $\vec{x}$ while the first three moments of $\vec{x}$ are related to the mean, fluctuation and skewness of the probability distribution:

$$N_C = \sum_{k=1}^{\infty} x_k, \qquad n_C = \sum_{k=1}^{\infty} k x_k / \sum_{k=1}^{\infty} x_k$$

$$<N> = \sum_{k=1}^{\infty} k x_k, \qquad <(N-<N>)^m> = \sum_{k=1}^{\infty} k^m x_k \tag{20}$$

The last result is only valid for $m = 2, 3$ with the value $m = 2$ giving the variances and $m = 3$ gives the skewness. Since $\log Z_{gc} = N_C = \Sigma x_k$, the $x_k$ represents the contribution to total $N_C$ coming from a cluster or clan of size $k$. The clan variables can also be related to the void variables. The connections are

$$N_C = Ln[Z_{GC}(\vec{x}, u = 1)] = -Ln P_0$$

$$\chi \equiv -Ln P_0 / <N> = N_C / <N> = 1/n_C \tag{21}$$

The last result shows that the void probability and moments of the $x_k$ distribution are connected. In the next three subsections, three examples will be given starting with the familiar negative binomial result but now done in the framework of a statistical model.

4.4 *Negative binomial case*

The $x_k$ that leads to a negative binomial distribution is $x_k = xt^k/k$ which has

$$Z_{gc} = (1-tu)^{-x} = 1 - \binom{-x}{+1} tu + \binom{-x}{+2}(tu)^2 - \ldots = 1 + \binom{x}{x-1} tu + \binom{x+1}{x-1}(tu)^2 + \ldots$$

$$Z_N = \frac{x(x+1)(x+2)\ldots(x+N-1)}{N!} t^N = \frac{\Gamma(x+N)}{N!\Gamma(x)} t^A = \binom{x+N-1}{N} t^N \tag{22}$$

The ratio $Z_N / Z_{N+1} = N/((N+x)t) \to 1/t$ as $N \to \infty$ with $x$ fixed. The $P_N$ is

$$P_N = \binom{x+N-1}{N} t^N (1-t)^x \tag{23}$$

The $<N>$ of the grand canonical ensemble is $<N>=xt/(1-t)$ at $u=1$. Solving for $t$ gives $t=(<N>/x)/(1+<N>/x)$ which when substituted into eq.23 gives the negative binomial distribution of eq.5. The variance $\delta N^2 = xt/(1-t)^2 =<N>(1+<N>/x)$. The clan variables $N_C, n_C$ for the negative binomial are

$$N_C = -xLn(1-t) = xLn(1+<N>/x)$$
$$n_C = -\frac{t}{(1-t)Ln(1-t)} = \frac{<N>/x}{Ln(1+<N>/x)} = \frac{1}{\chi} \qquad (24)$$

*4.5 Hybrid model; signal/noise or coherent/chaotic model*

A hybrid model, also used in quantum optics (Klauder and Sudarshan 1968), extrapolates between two limits, one being a Poisson distribution, the other being a negative binomial distribution. In quantum optics, the Poisson distribution arises from a coherent photon source while the negative binomial characterizes a chaotic source. To simulate this type of behavior, $x_k$ is taken as a mixture (Mekjian 2001) $x_k = yt^k + xt^k/k$. The $Z_{gc}$ and $Z_N$ for this $x_k$ are

$$Z_{gc} = \frac{1}{(1-t)^x} \exp(y\frac{t}{1-t})$$
$$Z_N = (-1)^{N+1} U[-N+1, x, -y] \frac{t^N}{N!} = NyL^1_{N-1}(-x)t^N \qquad (25)$$

with $U[-N+1, x, -y]$ a confluent hypergeometric function and $L^1_{N-1}(-x)$ an associated Laguerre polynomial. The $y, t$ are given by a coherent signal parameter $S$ and chaotic noise parameter $Nl$ through the relations

$$t = \frac{Nl/x}{1+(Nl/x)},$$
$$y = S\frac{1}{(Nl/x)(1+(Nl/x))} \qquad (26)$$

The $<N>= S + Nl$ and

$$<N^2>-<N>^2 = <N>+<N>^2 \frac{Nl}{Nl+S}(1+\frac{S}{1+(Nl/x)})\frac{1}{x} \qquad (27)$$

If $S = 0$, $x_k = xt^k/k$ and a negative binomial distribution is obtained. If $Nl \to 0$,

$x_k \to S\delta_{k,1}$ and only $x_1 \neq 0$ so that a Poisson distribution results.

4.6 *Thermodynamic model*

Another choice for $x_k$ that leads to simple results which is important for this paper is:

$$x_k = \frac{x}{k}\binom{2k-1}{k-1}\frac{t^k}{2^{2(k-1)}} \tag{28}$$

This case will be referred to as the thermodynamic model since some of the results that will be developed here are contained in a previous model, initially developed by Saslaw and Hamilton 1984 and further studied and modified by Fry 1986, in the context of galaxy clustering. The initial Saslaw and Hamilton result gives a probability for finding $N$ galaxies

$$P_N(<N>,b) = \frac{<N>(1-b)}{N!}[<N>(1-b)+Nb]^{N-1}\exp[-<N>(1-b)-nb] \tag{29}$$

The $<N> = \Sigma N P_N(<N>,b)$, while the fluctuation and skewness are given by

$$\begin{aligned}&<(N-<N>)^2> = <N>/(1-b)^2\\&<(N-<N>)^3> = <N>(1+2b)/(1-b)^4\end{aligned} \tag{30}$$

The fluctuation can then used to obtain the connection $1/(1-b)^2 = (1+\xi_2 <N>)$ and the void scaling properties of $-LnP_0/<N> = 1/(1+<N>\xi_2)^{1/2}$. Fry later modified the void distribution $P_0(<N>,b) = \exp(-<N>(1-b)) = \exp(-<N>/(1+\xi_2 <N>)^{1/2})$ into the distribution $P_0(<N>,b) = \exp(-\xi_2^{-1}[(1+\xi_2 <N>)^{1/2}-])$. Fry pointed out that the original Saslaw/Hamilton distribution is not a discrete realization of a continuous background number density, but in the continuum limit of large $<N>$ it could be processed into one that is related to such a distribution. The approach developed in this paper gives a systematic method of obtaining such probability functions. As an example the Fry extension of the thermodynamic result follows from the $x_k$ given above. The corresponding probability density $P_N$ will be given below. As mentioned, this result will be shown to be a special case of a more general result which also includes the popular negative binomial model.

As already pointed out, the probability associated with this model for $x_k$ also appeared in quantum optics in a model initially developed by Glauber 1963. The photon count probability arises when considering field emission from Lorentzian line shapes. The reformulation of the model in terms of the statistical model discussed here can be found in Mekjian (2001,2002). In this reference, the model was called the Lorentzian-Catalan model or *LC* model. The Catalan numbers $Cl_k \equiv (1/k)\,[(2(k-1))!/((k-1)!(k-1)!)]$

appear in eq.28. Using Stirlings approximation, the large $k$ behavior of $Cl_k \sim 2^{2(k-1)}/\sqrt{\pi}k^{3/2}$. This combinatoric factor counts the number of diagrams associated with the evolution of a process when stochastic probability variables are included-see sect.4.7. The associated $Z_N$ is

$$Z_N = (2x)^{2N} U[N,2N,4x]\frac{t^N}{N!} = \frac{(2x)^{2N} t^N}{N!}\left(\frac{4^{N/2}\exp(2x)x^{1/2-N}}{\sqrt{\pi}} K_{N-1/2}(2x)\right) \quad (31)$$

and the $Z_{gc}(u) = \exp[2x(1-\sqrt{1-tu})]$. The $U$ is a confluent hypergeometric function, while $K_{N-1/2}$ is a Bessel $K$ function of fractional order $N-1/2$. The ratio:

$$\frac{U[N,2N,4x]/N!}{U[N+1,2(N+1),4x]/(A+!)!} \to 4x^2 \quad (32)$$

as $N \to \infty$ from properties of the confluent hypergeometric function. Using this result the ratio of canonical partition functions scale as:

$$\frac{Z_N}{Z_{N+1}} \to \frac{1}{t} \quad (33)$$

as $N \to \infty$. This last ratio is independent of $x$. The probability of $N$ is just $P_N = Z_N / Z_{gc}(u)_{u\to 1}$. The $<N>$ and $\delta N^2$ are

$$<N> = \frac{xt}{\sqrt{1-t}}$$
$$\delta N^2 = \frac{(2-2t+t)xt}{2(1-t)^{3/2}} = <N> + \frac{<N>^2}{2x\sqrt{1-t}} \quad (34)$$

Thus $\xi_2 = 1/(2x\sqrt{1-t})$ and $\xi_2$ depends not only on $x$ but also on $t$. The quantity $\xi_2 <N> = t/(2(1-t))$ is x independent. The clan variables for the thermodynamic model are

$$N_C = 2x(1-\sqrt{1-t})$$
$$n_C = \frac{t}{2(\sqrt{1-t})(1-\sqrt{1-t})} \quad (35)$$

The asymptotic $1/k^{3/2}$ dependence of the thermodynamic model suggests a close correspondence with Bose-Einstein phenomena in 1 dimension. In $d$ dimensions, the $x_k = (L^d/\lambda_T^d)/k^{1+d/2}$ as given in sec. 4.2 and for $d=1$, $x_k = (L^1/\lambda_T)/k^{1+1/2} \sim 1/k^{3/2}$.

4.7 *Stochastic variables*

A stochastic probability variable can be incorporated into statistical models. Replacing $x, t$ with

$$x = \beta_C / 4p$$
$$t = 4p(1-p) \tag{36}$$

the factor $xt^k \rightarrow \beta_C p^{k-1}(1-p)^k \, 2^{2(k-1)}$. The commonly occurring factors $(1-t)$ and $\sqrt{1-t}$ are $(1-t) = (1-2p)^2$ and $\sqrt{1-2p} = \sqrt{(1-2p)^2} = /1-2p/$. The $/1-2p/$ is the absolute value of $1-2p$; for $p \leq 1/2$, the $/1-2p/ = 1-2p$ and for $p \geq 1/2$, the $/1-2p/ = 2p-1$. The $p$ can represent the probability that a cluster of galaxies will grow in size by one, while the factor $(1-p)$ is the probability that this additional growth survives without further increase or accretion. The $p$ is then a gravitational coalescence parameter or adhesion parameter in a statistical approach. Different adhesion approximations have been incorporated into dynamical models such as in the Zel'dovich (1970) picture. Sahni *et al.* (1994) studied the dynamical evolution of voids in the framework of an adhesion model.

If this change of variables is incorporated into the $x_k$ of eq.28, the following expression for $x_k$ results.

$$x_k = \frac{x}{k}\binom{2k-1}{k-1}\frac{t^k}{2^{2(k-1)}} = = \beta_C \frac{1}{k}\binom{2k-1}{k-1}p^{k-1}(1-p)^k \tag{37}$$

The Catalan combinatoric factor $Cl_k = (1/k)\,[(2(k-1))!/((k-1)!)^2]$ counts the number of diagrams with $(k-1)$ factors of $p$ and $k$ surviving lines each with probability $1-p$. For $k = 1,2,3,4$ the Catalan numbers are $Cl_1 = 1, Cl_2 = 1, Cl_3 = 2, Cl_4 = 5$. The $Cl_1 = 1$ represents a single line which survives with probability $1-p$ and has no adhesive junction. Consequently, $x_1 = 1 p^0 (1-p)^1$. The $Cl_2 = 1$ represents the single diagram with an incoming line which splits into two with probability $p$ at the junction. Each of the two lines formed survive with probability $(1-p)$ giving the $x_2 = 1p(1-p)^2$. The $Cl_3 = 2$ represents the two diagrams that can be obtained from the previous case of $Cl_2 = 1$ by splitting either line with probability $p$ at the junction. Thus $x_3 = 2p^2(1-p)^3$ since two $p$'s occur and three lines that are generated now survive each with probability $(1-p)$. The $Cl_4 = 5$ will have 5 possible diagrams associated with three $p$ factors and four $(1-p)$ factors. The $Cl_4 = 5$ can be generated from the $Cl_3 = 2$ in a manner similar to obtaining $Cl_3 = 2$ from $Cl_2 = 1$. The number of diagrams associated with $Cl_k$ grows exponentiall fast: $Cl_k = 2^{2(k-1)}/(k^{3/2}\sqrt{\pi})$ for large $k$ using Stirlings approximation. The $Z_N$ is

$$Z_N = \frac{1}{N!}(\beta_C(1-p))^N \sqrt{\frac{\beta_C}{2p}} \exp(\beta_C/2p) K_{N-1/2}(\beta_C/2p)\sqrt{\frac{2}{\pi}} =$$

$$\frac{1}{N!}\left(\frac{\beta_C}{p}\right)^{2N} (p(1-p))^N U[N, 2N, (\beta_C/p)] \tag{38}$$

and the $Z_{gc}$ (at $u = 1$) is determined by

$$Z_{gc} = \exp[2\frac{\beta_C}{4p}\{1-\sqrt{1-4p(1-p)}\}] = \exp[2\frac{\beta_C}{4p}\{1-/1-2p/\}] \tag{39}$$

with

$$Z_{gc} = \exp(\beta_C), \quad p \leq 1/2 ;$$
$$Z_{gc} = \exp(\beta_C(1-p)/p), \quad p \geq 1/2 \tag{40}$$

The $Z_{gc}$ is independent of $p$ for $p \leq 1/2$, but depends on $p$ for $p \geq 1/2$. A further discussion of this point will be presented in sec.8. The $<N>$ and $\delta N^2$ are

$$<N> = \beta_C \frac{1-p}{/1-2p/} \qquad \delta N^2 = <N>(1+\frac{2p}{/1-2p/}<N>) \tag{41}$$

The $<N>$ involves $\beta_C$ and $p$. The clan variable $N_C$, for $p \leq 1/2$, is $N_C = \beta_C = \Sigma x_k$. The $x_k$ is the contribution to $N_C$ from a clan of size $k$. While each $x_k$ is determined by $p$ and $\beta_C$, the sum of $x_k$ over $k = 1,2,...\infty$ is independent of $p$ for $p \leq 1/2$. The mean number of particles per clan $n_C = <N>/N_C = (1-p)/1-2p/^{(-1)}$ for $p \leq 1/2$. When $p \to 1/2$, $<N> \to \infty$ and $n_C \to \infty$, $1/n_C \to 0$. The $x_k$ at $p = 1/2$ falls as a pure scale invariant power law which, for large $k$, is $x_k \sim \beta_C/2\sqrt{\pi}k^{3/2}$. Thus, the first moment of $x_k$ with respect to $k$ and all higher moments diverge at $p = 1/2$ or $t = 1$.

4.8 *Generalized hypergeometric model; the parameter $a$ and Levy index $\alpha$, Fisher exponent $\tau$*

The results of two previous examples, binomial and thermodynamic cases, can be united into one model by introducing a variable $a$ into $x_k$. Specifically, the $x_k$ is taken to be of the form

$$x_k = x \frac{[a]_{k-1}}{k!} t^k \tag{42}$$

with $[a]_m = a(a+1)...(a+m-1) = \Gamma(a+m)/\Gamma(a)$ and $[a]_0 = 1$. At the special value $a = 1$ the $x_k$ is

$$x_k = \frac{x}{k} t^k, \qquad a = 1 \qquad (43)$$

which is the negative binomial result, At $a = 1/2$ the $x_k$ is

$$x_k = \frac{x}{k}\binom{2k-2}{k-1}\frac{t^k}{2^{2(k-1)}}, \qquad a = \frac{1}{2} \qquad (44)$$

which is the thermodynamic model result. For large $k$, the general $x_k$ varies with $k$ as

$$x_k \sim x \frac{1}{k^{2-a}} t^k \qquad (45)$$

This $x_k$ can be used in cluster distributions (Lee and Mekjian 2004) where the Fisher exponent $\tau$ and the parameter $a$ are related by $\tau = 2 - a$.

For this generalized $x_k$ model, the $<N>$ and $\delta N^2$ are

$$<N> = xt/(1-t)^a$$
$$\delta N^2 = \frac{xt}{(1-t)^{1+a}}(1+(-1+a)t) = <N> + \frac{a<N>^2}{x(1-t)^{1-a}} \qquad (46)$$

This last result gives $\xi_2 = a/(x(1-t)^{1-a})$ for the coefficient in front of $<N>^2$. The choice $a = 1$ gives the familiar and simple negative binomial result $\xi_2 = 1/x$. The choice $a = 1/2$ has $\xi_2 = 1/(2x\sqrt{1-t})$, the thermodynamic result.

The logarithm of the grand canonical partition function $Z_{gc} (= \Sigma x_k u^k)$ is

$$Ln(Z_{gc}) = (xtu)_2F_1(a,1;2;tu) = x(1-(1-t)^{1-a})/(1-a) = N_C \qquad (47)$$

where $_2F_1(a,1;2;tu)$ is a hypergeometric function. The underlying partition weight $W_N(\vec{n},\vec{x})$ is obtained by substituting eq.28 into eq.18. The partition function $Z_N$ can be obtained from the recurrence relation of eq.16 using the $x_k$ of eq.42.

In general, the hypergeometric function $_2F_1(a,b;c;z)$ is

$$_2F_1(a,b;c;z) = \sum_{m=0}^{\infty} \frac{[a]_m [b]_m}{[c]_m m!} z^m \qquad (48)$$

The hypergeometric model that was just considered has $b = 1$ and $c = 2$. More general cases can also be developed by allowing $b,c$ to have other values. The case $b = 1$ and $c = 2$, with $a$ in the range $0 \leq a \leq 1$ have associated probability distribution which are stable for $t < 1$. Probability distribution of his type appear in mathematics as stable Levy distributions which have an index $\alpha$, with $1 \geq \alpha \geq 0$. The Levy index $\alpha$ is connected to $a$, for $1 \geq a \geq 0$, by the simple relationship $1 - \alpha = a$. A negative binomial will have $a = 1, \alpha = 0$ while the case $a = 1/2$ will also have $\alpha = 1/2$. The Poisson limit is $a \to 0$, $\alpha \to 1$. At $a = 0$ only $x_1$ is non-zero.

Many cluster distributions and scale free power laws have $\tau$ in the range $2 < \tau < 3$. This leads to a faster fall off than $1/k^{2-a} = 1/k^{1+\alpha}$. The hypergeometric model $_2F_1(a,1;3,z)$ with $b = 1$ and $c = 3$ has $Ln(Z_{gc}) = (xtu)_2F_1(a,1;3;tu) = \Sigma x_k u^k$ with $x_k \sim xt^k / k^{3-a}$ for large $k$. For $a = 1/2$, the $x_k \sim xt^k / k^{5/2}$. Thus $\Sigma x_k$ and $\Sigma k x_k$ are now both finite. The results for $a = 1/2, b = 1, c = 3$ approximates the behavior of Bose-Einstein condensation for particles in a 3 dimensional volume where $x_k = x/k^{5/2}$ as given in sec.4.2. The zeta functions that appear in exact theory, such as
$\varsigma(5/2) = \Sigma 1/k^{5/2} = 1.341, \varsigma(3/2) = \Sigma 1/k^{3/2} = 2.612$ are replaced by 4/3=1.333 and 8/3=2.667, respectively. Further discussions of these result will be given elsewhere.

## 5. VOID SCALING RELATIONS AND HIERARCHICAL STRUCTURE FUNCTIONS

5.1 *Void probability and void scaling relations*

This section contains a more detailed discussion of void scaling relations in subsect. 5.1 and hierarchical structure functions in subsect. 5.2. The void probability is $P_0$ and, in a statistical model, the $P_0$ is determined by

$$P_0 = Z_0 / Z_{gc}(u=1) = 1/Z_{gc}(u=1) = \exp(-N_C) \tag{49}$$

The void function $\chi$ has several useful forms given by

$$\chi \equiv -Ln(P_0)/<N> = Ln(Z_{gc})/<N> = N_C/<N> = n_C. \tag{50}$$

The scaling behavior of $\chi$ relates $\chi$ to a function associated with the variance written as $<N^2> - <N>^2 \equiv <N> + \xi_2 <N>^2 = <N>(1 + \xi_2 <N>)$. The quantity of interest is the function $\xi_2 <N> = (<N^2> - <N>^2 - <N>)/<N>$ which is the ratio of the second factorial moment to the first factorial moment, with the later being $<N>$. The $\xi_2 <N>^2$ term gives the departures from Poisson statistics and $1 + \xi_2 <N>$ is an enhancement factor. For a negative binomial distribution the $\chi$ is

$$\chi_{NB} = \frac{1}{\xi_2 <N>} Ln[1+\xi_2 <N>] \qquad (51)$$

The corresponding quantity in the thermodynamic model is labeled $\chi_{LC}$ and is

$$\chi_{LC} = \frac{1}{\xi_2 <N>}[\sqrt{(1+2\xi_2 <N>)}-1] = \frac{2}{\sqrt{(1+2\xi_2 <N>)}+1} \qquad (52)$$

The $\chi_{NB}$ and $\chi_{LC}$ are special cases of the hypergeometric model of sec.4.8 where $\chi_a$ is

$$\chi_a = \frac{1}{(1-a)(\xi_2 <N>/a)}[\left(1+\frac{\xi_2 <N>}{a}\right)^{1-a}-1] \qquad (53)$$

The $\chi_{NB} = \chi_{a=1}$ and $\chi_{LC} = \chi_{a=1/2}$. When $\xi_2 <N> \to \infty$, $\chi_a \to 0$. The vanishing of $\chi_a$ with $\xi_2 <N> \to \infty$ is linked to $(1/n_C) \to 0$ with $t \to 1$ or $p \to 1/2$. In turn $(1/n_C) \to 0$ is connected to the fact that $\Sigma k x_k \to \infty$ with $t \to 1$ or $p \to 1/2$. This result is a consequence of the asymptotic power law fall off of $x_k$ at $t=1$ or $p \to 1/2$, i.e. $x_k \sim 1/k^{2-a} = 1/k^\tau = 1/k^{1-\alpha}$. For $\Sigma k x_k \to \infty$, and for $\Sigma x_k$ to be finite, the variable $a$, or Fisher exponent $\tau = 2-a$ or Levy index $\alpha = 1-a$ lie in certain ranges. The exponent $\tau$ is in the range $1 < \tau \le 2$, the index $\alpha$ is in the range $0 \le \alpha < 1$ and the hypergeometric variable $a$ is in the range $0 \le a < 1$.

As shown at the end of sec.4.8 generalized hypergeometric models can also be constructed with $x_k$ falling with a higher power of $k$ such as $x_k \sim 1/k^{3-a} = 1/k^\tau$. If $3 > \tau > 2$, $<N> = \Sigma k x_k \sim \Sigma 1/k^{\tau-1}$ is finite and $n_C = <N>/N_C$ and $1/n_C$ are also finite. Then $\chi \to N_C/<N> = 1/n_C$, a non-zero constant, as $\xi_2 <N> \to \infty$. Thus, the vanishing of $\chi$ with $\xi_2 <N> \to \infty$ is an indication of $\tau$ in the range $1 < \tau \le 2$.

For small $\xi_2 <N>$ eq.53 gives

$$\chi_a = 1 - \xi_2 <N>/2 \qquad (54)$$

and $\chi_a$ is independent of $a$. The $\chi_a$ of eq.54 is referred to as the Gaussian approximation. The independence of $\chi_a$ with $a$ is a realization of the fact that all models can be replaced by a Gaussian approximation for small $\xi_2 <N>$. The Gaussian approximation can be realized by keeping only $x_1$ and $x_2$ terms. Then $Z_{gc} = \exp(x_1 + x_2)$, $<N> = x_1 + 2x_2$, and $<N^2> - <N>^2 = x_1 + 4x_2 = <N> + 2x_2 = <N> + \xi_2 <N>^2$. Thus, $2x_2 = \xi_2 <N>^2$. The void scaling function $\chi_2 \equiv -LnP_0/<N>$ is simply

$$\chi_2 \equiv (x_1 + x_2)/<N> = (<N> - x_2)/<N> = 1 - \xi_2 <N>/2 \qquad (55)$$

The hybrid coherent signal/chaotic noise model has $\chi_{SNl}$, $\xi_2$ and $\xi_2 <N>$ that are

$$\chi_{SNl} = \frac{x}{<N>} Ln[1 + \frac{Nl}{x}] + \frac{S}{<N>} \frac{1}{(1+(Nl/x))}$$
$$\xi_2 <N> = \frac{Nl}{x}(1 + \frac{S}{Nl+S}) = \frac{Nl}{x}(1 + \frac{S}{<N>}) \quad (56)$$
$$\xi_2 = \frac{1}{x}(1 - \frac{S}{<N>})(1 + \frac{S}{<N>})$$

For $Nl \to 0$, $\xi_2 \to 0$ and $\chi_{SNl} \to \chi_P = 1$ for any $\xi_2 <N>$ which is the Poisson behavior of $\chi$. For $S \to 0$, $\xi_2 \to 1/x$ and the negative binomial behavior results. For $S \neq 0, Nl \neq 0$, a void scaling behavior in $\xi_2 <N>$ does not exist. To further illustrate how the void scaling relation is violated by the presence of a Poisson term with strength $S$, the $\chi_{SNl}$ can be rewritten as

$$\chi_{SNl} = \frac{(1-(\frac{S}{<N>}))}{\frac{\xi_2 <N>}{1+S/<N>}} Ln[1 + \frac{\xi_2 <N>}{1+S/<N>}] + \frac{S}{<N>} \frac{1}{1 + \frac{\xi_2 <N>}{1+S/<N>}} \quad (57)$$

This last relation shows how the negative binomial scaling is violated by terms involving $S/<N> = S/(S+Nl)$ which is the relative signal strength to signal plus noise level.

5.2 *Hierarchical structure relations and reduced cumulants*

Another important aspect of the models considered here is their hierarchical structure. The factorial cumulants $f_q$ of order $q$ are obtained from an expansion of $Ln(Z_{gc}(\vec{x}, u))$:

$$Ln(Z_{gc}(\vec{x}, u)) = \Sigma(u-1)^q f_q / q! = \Sigma x_k u^k \quad (58)$$

The $f_q$ and $x_k$ are connected by

$$f_q = q! \sum_{n=q}^{\infty} \binom{n}{q} x_n = q! \sum_{n=1}^{\infty} \binom{n}{q} x_n = \sum_{n=1}^{\infty} n(n-1)...(n-q+1)x_n \quad (59)$$

The $f_2 = 2! \sum_{n=2} n(n-1)x_n / 2 = <A^2> - <A>^2 - <A>$. For a Poisson distribution $f_2 = 0$ and all higher $f_q = 0$, since only $x_1 = 0$. For a hypergeometric model with parameter $a$

$$f_q = \frac{\Gamma(a+q-1)}{\Gamma(a)} \frac{xt^{q-1}}{(1-t)^{a+q-1}} \tag{60}$$

The normalized factorial cumulant $k_q = f_q / <N>^q$ satisfies a relation which connects $k_q$ to $k_2$ which reads

$$k_q = A_q k_2^{q-1} \tag{61}$$

The $k_2 = \xi_2$ since $f_2 = <N^2> - <N>^2 - <N> = \xi_2 <N>^2$. The coefficient $A_q$ is given by

$$A_q(a) = \frac{\Gamma(a+q-1)}{\Gamma(a) a^{q-1}} \tag{62}$$

The $A_2(a) = 1$ since $\Gamma(a+1) = a\Gamma(a)$. For the negative binomial $A_q(a=1) = (q-1)!$. The existence of relations of the form $k_q = A_q k_2^{q-1}$ relating $k_q$ to powers of $k_2$ express an hierarchical structure on a reduced cumulant level. The above result shows that the hierarchical structure is still preserved for the more general hypergeometric case with variable exponent $a$. The thermodynamic model has $a = 1/2$ which when substituted into $A_q(a)$ gives $A_q(1/2) = 2^{q-1} \Gamma(1/2 + q - 1) / \Gamma(1/2) = (2q-3)(2q-5)(2q-7)...1$. For $q = 3$ (skewness) and $a = 1/2$, the $A_3(1/2) = 4\Gamma(5/2)/\Gamma(1/2) = 3$. For a negative binomial $a = 1$ and the skewness $A_3(1) = 2$. The $A_q(a)$ increases with decreasing $a$ with $q$ held constant.

### 6. COSMIC STRINGS

As noted in the introduction, understanding the origins of large scale structure and anisotropy of the cosmic microwave background has proceeded along several lines. One of these approaches is based on symmetry breaking during a phase transition in the early universe which plays an important role in its evolution in time. Associated with these phase transitions are topological defects such as cosmic strings, domain walls, and texture. Loops of cosmic string may be centers for the accretion of galactic matter and can thus be a source of the clumping of matter seen in the large scale structure of the universe. However, if stings continually intersect and break into smaller strings, their effect is reduced and in fact they might disappear even before they have a chance to accumulate matter that forms galaxies. A simple model was proposed by Smith and Vilenkin (1987) to study this feature. Here, a brief summary will be given since some of its formal features appeared in a statistical model when stochastic variables were introduced. Moreover, this formal similarity with results from cosmic strings will be helpful in the next section which discusses a further parallel with percolation theory.

The Smith/Vilenkin model is based on an initial ancestral string which intersects itself and generates daughter strings in the process of intersection. A probability of intersection is $p$ and the probability of survival without further intersection is $1-p$. Daughter or descendent loops follow the same process with the same probabilities $p$ and $1-p$. The probability of having $n$ loops was given as

$$\widehat{P}_n = A_n p^{n-1}(1-p)^n \tag{63}$$

where $A_n$ satisfies the recurrence relation $A_n = \Sigma A_k A_{n-k}$, where the sum runs from $k=1$ to $n-1$ and $A_1 = 1$. This recurrence relation follows from the observation that the first daughter of the ancestral loop can produces $k = 1,2,...,n-1$ off spring while the second daughter itself produces $n-k$ descendents. The solution to the recurrence relation is

$$A_n = \frac{1}{n}\binom{2n-2}{n-1} \tag{64}$$

For large $n$

$$\widehat{P}_n = \frac{1}{n}\binom{2n-2}{n-1} p^{n-1}(1-p)^n \approx \frac{1}{4\sqrt{\pi p}} \frac{(4p(1-p))^n}{n^{3/2}} \tag{65}$$

For $p \geq 1/2$ an $\infty$ cascade occurs, while for $p < 1/2$ a finite number of loops are produced. The probability $\mathcal{P}$ that an $\infty$ fragmentation happens satisfies the condition

$$\mathcal{P} + \sum_{n=1}^{\infty} \widehat{P}_n = 1 \tag{66}$$

and thus $\mathcal{P} = 0$ for $0 \leq p \leq 1/2$, and $\mathcal{P} = (2p-1)/p$ when $1/2 \leq p \leq 1$. The mean number of loops generated is $<n> = (1-p)/(1-2p)$ and the variance $\delta n^2 = p(1-p)/(1-2p)^3$. The $\widehat{P}_n$ has some features that resemble some elements that appear in the thermodynamic model. Specifically, the $x_k$ of eq.37 has a similar structure in the stochastic variable $p$ and associated $1-p$ and in the combinatoric factor $A_n$. The $x_k$ of eq.37 has the additional factor $\beta_C$ which is the number of clans associated with the clan variable description. Also the $\widehat{P}_n$ starts at $n=1$. The probability that is generated by the $x_k$ is, for $0 \leq p \leq 1/2$, given by

$$P_N = \frac{1}{N!}\left(\frac{\beta_C}{p}\right)^{2N} (p(1-p))^N U[N, 2N, (\beta_C/p)] \exp[\beta_C] \tag{67}$$

Comparing the thermodynamic model $P_N$ with the cosmic string $\widehat{P}_n$ for $n = N \gg \beta_C$ and for $p \to 1/2^-$, the ratio

$$\frac{P_N}{\widehat{P}_{n=N}} \to \beta_C \tag{68}$$

or, for $N \gg \beta_C$ and $p \to 1/2^-$

$$P_N \to \approx \beta_C \frac{1}{4\sqrt{\pi}p} \frac{(4p(1-p))^N}{N^{3/2}} \to \beta_C \frac{1}{2\sqrt{\pi}} \frac{1}{N^{3/2}} \tag{69}$$

These results shown that the thermodynamic model has a probability distribution for large $N$ and $p \to 1/2$ that is a scaled version of the cosmic string probability, with the scaling factor the number of clans $\beta_C$. The two distributions differ for small $N$. For $\beta_C = 1$, the two distributions are asymptotically the same, but still differ for small $N$ because of the void probability $P_0$. The $P_N < \widehat{P}_{n=N}$ for $\beta_C = 1$ and $p = 1/2$. The $P_N$ falls as a pure power law $1/N^{3/2}$ and is thus scale free as $p \to 1/2$. For $p < 1/2$, a decreasing exponential term in $N$ is present in $P_N$, i.e. the term involving $p$ as $(4p(1-p))^N = \exp(-/\lambda_p / N)$ with $\lambda_p = \ln(4p(1-p)) < 0$ for $p \neq 1/2$. The $P_N$'s are no longer scale free for $p \neq 1/2$. At $p = 1/2$, the $<N> \to \infty$ since $P_N \to 1/N^{3/2}$. The $\exp(-/\lambda_p / N)$ stabilizes the distribution for $p \neq 1/2$. An infinite $<N>$ is characteristic of a Levy unstable distribution. An unstable Levy distribution can also be made stable by truncating $P_N$: $P_N = 0$, for $N > N_C$, where $N_C$ is a cut-off $N$.

## 7. POSSIBLE PARALLELS WITH PERCOLATION THEORY AND PHASE TRANSITIONS

The region $p > 1/2$ in the thermodynamic model with stochastic variables corresponds to a new branch were the number of clans changes from $N_C = \beta_C$, a constant for all $p \leq 1/2$, to a $p$ dependent quantity. This sudden change suggests an ansatz where an addition quantity is introduced to keep $N_C$ a constant. This section explores this possibility and its consequences. Further investigations are in progress and the results p[resent here are tentative. First, some motivation is given for this possibility. In percolation, when the bond or site probability crosses the critical $p_c$ an infinite cluster appears with strength $S$ which depends on $p$. For percolation on a Bethe lattice or Cayley tree this $S = p\{1 - [(1-p)/p]^3\}$. In Bose-Einstein condensation, when the temperature falls below $T_C$, the ground state occupancy is singled out and treated separately. Then, suddenly, at $T_C$ the ground state starts to be filled with a macroscopic number of particles. In the cosmic string discussion given above, for $p > 1/2$, a $\mathcal{P}$ suddenly appeared which represented the fact that the splitting of the ancestral loop and descendents now goes on forever. The $\mathcal{P} = (2p-1)/p$. The extra amount needed to keep the clan number fixed at $N_C = \beta_C$ for $p > 1/2$ is labeled $\varphi_C$ and this $\varphi_C = \Theta(p - 1/2)\, \beta_C(2p-1)/p$, where $\Theta(p-1/2)=1$ for $p>1/2$ and $\Theta(p-1/2)=0$ for $p<1/2$. The additional $\varphi_C$ is part of the $\Sigma x_k = N_C$, such that the sum is now fixed at the constant $\beta_C$ for all $0 \leq p \leq 1$. The $\varphi_C$ is only defined by its total sum rule and not by a specific distribution in $x_k$. To proceed, a modified form of a signal/noise or Poisson/negative binomial model (sec. 4.5) will be used as a possible option. The Poisson signal level will be $\varphi_C$ and the thermodynamic model will now replace the negative binomial. The $\varphi_C$ is then an additional Poisson piece and the total $\varphi_C$ strength is added to $x_1$: $x_1 \to (1-p) + \varphi_C = (1-p) + \Theta(p-1/2)\beta_C(2p-1)/p$. The $x_k$, for $k \neq 1$, is unchanged and is therefore $x_k = Cl_k p^{k-1}(1-p)^k$. For $p \leq 1/2$ the additional Poisson amplitude vaninhes while for $p \to 1$, only a pure Poisson distribution survives. The $<N>$ is now

$$<N> = \varphi_C + \beta_C \frac{1-p}{/2p-1/} \tag{70}$$

and

$$<N^2> - <N>^2 = \varphi_C + \beta_C \frac{(1-p)(1-2p+2p^2)}{/2p-1/^3} \tag{71}$$

Then, as before, for $p \leq 1/2$, the additional Poisson term vanishes and

$$\chi = 1 - \frac{p}{1-p} = \frac{1}{\xi_2 <N>}[\sqrt{(1+2\xi_2 <N>)} - 1] = \frac{2}{\sqrt{1+2\xi_2 <N>}+1} \tag{72}$$

where $\xi_2 = 2p/(\beta_C(1-2p))$ and $\xi_2 <N> = 2p(1-p)/(1-2p)^2$. The $p$ in terms of $r \equiv 2\xi_2 <N>$ is $p = (1/2) - 1/(2\sqrt{1+r})$ and $p/(1-p) = (\sqrt{1+r}-1)/(\sqrt{1+r}-1)$. However, for $p > 1/2$, a scaling relation for $\chi$ in terms of the scaling variable $\xi_2 <N>$ is no longer valid because of the presence of the Poisson part $\varphi_C$. The $\chi$ now has a contribution from $\varphi_C$ and now reads ($p > 1/2$)

$$\chi = \frac{\beta_C}{\varphi_C + \beta_C \frac{1-p}{/2p-1/}} = \frac{1}{\hat{\varphi}_C + \frac{1-p}{/2p-1/}} \tag{73}$$

where $\varphi_C = \beta_C \hat{\varphi}_C$. The count distribution follows from the recurrence relation of eq.16.

## 8. CONCLUSIONS AND SUMMARY

A statistical approach was introduced which can be used to discuss various features associated with the distribution of galaxies and the large scale structure of the universe. These features include the clumping of galaxies, the associated production of voids and an hierarchical structure of correlations. Within the statistical framework, a generalized model was considered which contains a continuous parameter or variable called $a$. The parameter $a$ is related to the Levy stable index $\alpha$, $(1-a=\alpha)$, in probability distributions, and the Fisher critical exponent $\tau$ in cluster phenomena, ($\tau = 2-a$). The Fisher index $\tau$ describes the power law fall off of clusters sizes or clumps of matter at a phase transition or critical opalescence point. The Levy index characterizes the behavior of non-Gaussian probability distributions that have power law behaviors. The continuous parameter $a$, exponent $\tau$, or index $\alpha$ can be obtained from properties of a void scaling function. Namely, the value of $a, \alpha,$ or $\tau$ governs the asymptotic behavior of $\chi$ in terms of the scaling variable $\xi_2 <N>$. Moreover, a vanishing $\chi$ with $\xi_2 <N>$. was shown to exclude values of $\tau > 2$. For $\tau > 2$, $\chi \to$ a non-zero value as $\xi_2 <N> \to \infty$. Special values of $a$ give a void and hierarchical structure of frequently used phenomenological models for galaxy count distributions in cosmology. Specifically, the value $a=1$ gives the negative binomial distribution, while $a=1/2$ gives the distribution based on a thermodynamic model. As $a \to 0$, a random uncorrelated Poisson limit is realized in the galaxy count distribution. Results based on combinations of Poisson plus chaotic distributions such as the negative binomial distribution were also considered. How the Poisson part affects the hierarchical scaling behavior associated with the negative binomial part was studied in detail. Parallels were drawn with Bose-Einstein statistical correlations which lead to clumping of photons in phase space. These parallels with Bose-Einstein phenomena were also used to motivate the form of the statistical model that was developed. The Van Hove clan description of clumps and voids was incorporated into the description and connected to features of the statistical model.

A formulation of the statistical model in terms of stochastic probability variables was also given. Namely, a probability $p$ was incorporated into the statistical picture where $p$ is the probability that a galaxy adheres to another galaxy or group of galaxies to form a larger cluster of galaxies. A factor $1-p$ in the description represents the probability that a cluster of galaxies survives without further adhesive processes. Thus, an evolutionary growth process is incorporated into this phenomenological description of galaxy structure. A parallel was drawn with percolation phenomena where $p$ represents either the bond probability of adjacent points on a lattice or the site probability of a point on the lattice. For percolation on a lattice, when $p \geq p_c$, an infinite cluster appears where $p_c$ is a critical value of $p$. At $p = p_C$ a pure scale invariant power law in cluster sizes manifestes itself. For a thermodynamic model of galaxy clustering, the behavior of the distribution is also quite different for $p \leq 1/2$ and $p > 1/2$ and parallels the percolation theory result. Namely, for $p < 1/2$, the adhesion is terminated and the galaxy count distribution decays exponentially fast with size of the galaxy cluster . For $p \geq 1/2$, the adhesion can go on forever and an infinite supercluster is formed. For $p = 1/2$, the probability distribution becomes a scale invariant power law. This model also has some formal results that are similar to cosmic string results which were briefly mentioned.

## 8. ACKNOWLEDGEMENTS

This work was supported in part by a grant from the department of energy under DOE grant number DE-FG02-96ER-40987